**This is a preliminary version of the JASIST submission**

# Upper Tag Ontology (UTO) For Integrating Social Tagging Data

**[1]Ying Ding (dingying@indiana.edu)**

**Elin K. Jacob (ejacob@indiana.edu)**

*School of Library and Information Science, Indiana University, 1320 E 10th, Bloomington IN 47405 USA*
*Tel: +1 812 855 5388; Fax: +1 812 855 6166*

**Michael Fried (michael.fried@sti2.at)**
**Ioan Toma (ioan.toma@sti2.at)**

*Institute of Computer Science, University of Innsbruck , Technikerstraße 21a, 6020 Innsbruck, Austria*

**Erjia Yan (eyan@indiana.edu)**
*School of Library and Information Science, Indiana University, Bloomington, IN 47405 USA*

**Schubert Foo (sfoo@pmail.ntu.edu.sg)**

*Division of Information Studies, Nanyang Technological University, Singapore.*

**Abstract**

**Data integration and mediation have become central concerns of information technology over the past few decades. With the advent of the Web and the rapid increases in the amount of data and the number of Web documents and users, researchers have focused on enhancing the interoperability of data through the development of metadata schemes. Other researchers have looked to the wealth of metadata generated by bookmarking sites on the Social Web. While several existing ontologies capitalize on the semantics of metadata created by tagging activities, the Upper Tag Ontology (UTO) emphasizes the structure of tagging activities to facilitate modeling of tagging data and the integration of data from different bookmarking sites as well as the alignment of tagging ontologies. UTO is described and its utility in harvesting, modeling,**

[1] Corresponding author.

**integrating, searching and analyzing data is demonstrated with metadata harvested from three major social tagging systems (Delicious, Flickr and YouTube).**

## 1. Introduction

Initially, the World Wide Web (Web) was a syntactically structured platform, woven together by hyperlinks, which functioned primarily to provide access to read-only resources. More recently, the Web has emerged as an exciting multimedia world, but it provided little support for users to share and to collaborate. Now, instead of the one-way presentation of resources, the Web provides a new environment for social networking and information sharing. It has evolved from a place to read to the place to write and to share. The term “Social Web”, introduced by Peter Hoschka in 1998, emphasizes the function of the Web as a social medium. In this paper, the Social Web is extended to include any Web-related technology, phenomenon or development that enhances the social nature of the Web.

Data mediation and data integration have been central concerns of IT for decades (Batini, Lenzerini, & Navathe, 1986; Rahm & Bernstein, 2001). With the advent of the Web, interest in these issues has exploded due to the growing amount of data and numbers of resources and users on the Web. Currently, there is a focus on providing machine supported meditation on the Web (Antoniou & Harmelen, 2004; Berners-Lee, Hendler, & Lassila, 2001) through the medium of machine-processable metadata that has been added to resources.

There are several different methods for adding metadata to Web resources. The formal method, driven by the needs of the Semantic Web, involves construction of well-defined ontologies that are used as a framework for annotating resources. Unfortunately, this approach simply shifts the problem from the level of the data to the level of the ontology. Nevertheless, because ontologies capture shared understandings and conceptualizations in formal, machine-processable languages, the generation of some metadata statements can be automated within certain domains. Ontology generation, annotation and maintenance, however, are extremely time consuming and hardly scalable (Gomez-Perez, Fernandez-Lopez, & Corcho, 2003).

In contrast, the socially-driven approach, found in the social tagging sites of the Web 2.0 environment, allows users to contribute metadata for resources by tagging them with any term(s) they like, leading to the emergence of tag clouds and system folksonomies (Guo, Jacob & George, 2009). While this approach gives users the freedom to add personally relevant metadata to any resource, such metadata is not often well defined and may not reflect community consensus. Furthermore, it is not formally represented in a machine-readable way and data integration cannot be easily achieved usaing automatic methods (Mika, 2007).

The standards-based approach is yet another attempt to promote compatibility among systems, databases and services. Organizations such as the World Wide Web Consortium (W3C) have undertaken extensive efforts intended to specify, develop and deploy standards for the sharing of semantics. These efforts are a crucial step towards enhancing Web functionality and interoperability (Antoniou & Harmelen, 2004). A number of specialized metadata schemes have also been developed. For example, Friend of A Friend (FOAF) is a schema for representing relationships among people (http://www.foaf-project.org/); Simple Knowledge Organization System (SKOS) is a schema for creating thesauri, taxonomies and other knowledge representation schemes for use on the Web (http://www.w3.org/2004/02/skos/); Description of A Project (DOAP) is an XML/RDF schema for the description of open source software projects (http://trac.usefulinc.com/doap); Really Simple Syndication (RSS) is an XML-based metadata schema for news (http://Web.resource.org/rss/1.0/); Semantically-Interlinked Online Communities (SIOC) is a metadata schema intended to promote integration of information about online communities (http://sioc-project.org/); Dublin Core Terms (DCT) is a simple schema for representing resources in general and bibliographic resources in particular

(http://dublincore.org/2008/01/14/dcterms.rdf); and Gene Ontology (GO) is a schema that supports description of genes and gene attributes (http://www.geneontology.org/). There is also a growing trend to develop microformats, simple data formats that reuse existing standards to provide solutions for common problems (http://microformats.org/). For example, the Geo microformat (GEO) provides a simple structure for marking up geographic coordinates in XHTML, RSS or XML (http://www.ncbi.nlm.nih.gov/geo/). All of these represent efforts to alleviate the persistent problems of integrating data across systems.

This paper describes the Upper Tag Ontology, its utility in aligning existing social tagging ontologies and its application in harvesting, modeling, integrating, searching and analyzing tagging data from three major social tagging systems: Delicious, Flickr and YouTube. The paper is organized as follows: Section 2 provides background information about tags and social tagging; Section 3 presents current work on the development of tag ontologies; Section 4 describes the Upper Tag Ontology (UTO) and aligns its use in aligning existing social metadata standards; Section 5 explains the UTO crawler and how it was used to harvest tagging data from Delicious, Flickr and YouTube; and Section 6 describes the integration of tag data using UTO and presents three query scenarios. Using the integrated tag dataset, Section 7 identifies a core set of 1363 tags and demonstrates how the distribution of frequencies of tag assignment accords with Zipf's power law. Section 8 summarizes the usefulness of tagging data available on the Social Web and proposes future avenues for research with social tagging data.

## 2. Tags and Social Tagging

A tag is a keyword assigned by a user to represent the subject content, format, utility or affective characteristics of a bookmark, photograph, video, audio, post, wiki, blog or other online resources. The goal of tagging is to make a collection of resources easier to search, to discover, to share and to navigate. A tag does not have a formal semantic reference per se, but only the informal semantics attached by each individual. Furthermore, different users may have very different objectives when they add a tag to an online resource: some users may want to categorize online resources to make them easier to find in the future; some may want to share resources they have found with other users; some may want to review -- to vote on -- films, videos, or restaurants by offering their personal opinions; and some may want simply to reference or cite online documents. According to a survey undertaken by Pew Internet and American Life Project in December 2006, 28% of American internet users had tagged online resources such as photographs, blog posts, and news articles; and 7% of American internet users reported that they tagged online resources on a daily basis (Rainie, 2007).

Because different users frequently add dissimilar or even conflicting tags to the same resource, the act of tagging can be said to reflect the individuality and identity of each user. But tagging is also a social activity.  If the activities of individual taggers are viewed as a whole, social tagging reflects the collective behaviour of a community of users. In a user study of over 4,000 participants conducted by Suchanek, Vojnovic and Gunawardena (2008), analysis of 65,000 Delicious bookmarks indicated that the more popular a tag was, the more likely it was to be meaningful to users. Their analysis also provided support for the assumption that the more users who have tagged a resource, the more meaningful the more frequently assigned tags will be.

More importantly, because tags are socially generated metadata, they can reflect the collective intelligence of a community of taggers. The aggregate of all tags assigned within a social bookmarking system constitutes a system folksonomy (Guo et al., 2008). Because such a system folksonomy is comprised of semantically meaningful folksonomy networks -- aggregations of user generated tags within a topical domain (Guo et al., 2009) -- it can be interpreted as a lightweight ontology representing social agreement within groups of users. More importantly, because a system folksonomy is generated through an inductive or bottom-up approach to vocabulary creation, it speaks the same language as the users who have created it and makes resources easier to identify and to retrieve (Mathes, 2004). This stands in sharp contrast to traditional system vocabularies, where construction by experts generally follows a top-down approach. Unfortunately, when subject experts have full responsibility for representing a domain or application, the resulting vocabulary may overlook the needs and requirements of its end users.

One of the primary bottlenecks stalling the realization of the Semantic Web involves ontology-based annotation. In order to move Web 2.0 to the level of the Semantic Web, online resources must be annotated in accordance with recognized ontologies. To date, there have been two primary approaches to ontology annotation: automatic approaches and manual approaches. Automatic approaches are based on natural language processing technologies, but the precision of automatic approaches is still questionable in massive deployments. Manual approaches are often more effective but time-consuming and are not scalable to very large resource collections. Social tagging offers an alternative to automatic and manual approaches. Although user annotations of online resources can sometimes be dirty and/or noisy, a little semantics is better than nothing and can go a long way toward realization of the Semantic Web.

## 3. Tag Ontologies

As researchers became aware of social tagging, they began to investigate methods for using social tagging data and capitalizing on social tagging behaviors. One of the foremost pioneering efforts was undertaken by Tom Gruber, generally acknowledged in the Semantic Web community as having first described the value of ontologies for addressing problems of resource annotation and interoperability across systems (see Gruber, 1994). In 2005, Gruber proposed the use of an ontology to model tagging data and support collaborative filtering; he subsequently formalized a conceptual model of his tagging ontology (Gruber, 2007). Gruber's tagging ontology covered the basic elements of tagging activity (i.e., *object*, *tag*, *tagger,* and *source*), to which he added the notion of a "vote" (i.e., + or -).

Three other ontologies have also been developed for representing tagging data: the Social Semantic Cloud of Tags (SCOT) ontology (http://scot-project.org/), the Holygoat Tag Ontology (http://www.holygoat.co.uk/projects/tags/), and the Meaning Of A Tag (MOAT) ontology (http://moat-project.org/). In line with Gruber's conceptual model of a tag ontology, SCOT defines a core set of concepts and properties deemed necessary to represent the structure and semantics of a social tagging system. Core concepts in the SCOT ontology are *scot:Tagcloud*, *scot:Tag* and *scot:Coocurrence*. There are also 35 properties, including, for example, *scot:hasTag*, *scot:spellingVariant*, *scot:usedBy*, *scot:createdBy*, etc. To support interoperability and minimize redundancies across schemas, development of the SCOT ontology has been based on the concept of linked data (i.e., the reduction of barriers to connecting decentralized or previously unconnected but nonetheless related data records). SCOT reuses elements from several existing schemas, including SIOC, FOAF, SKOS, MOAT and DCT. For example, SCOT uses SIOC elements to describe site information and relationships among site resources, FOAF elements to represent a human or machine agent, SKOS elements to characterize the relationships among tags, and MOAT elements to define the meaning of an individual tag.

The Holygoat Tag Ontology, created by Richard Newman, models an instance of tagging as the reification of a relationship binding a tagger, a resource and a date to at least one tag. As with the SCOT ontology, the Holygoat Tag Ontology incorporates elements from existing ontologies in its definitions of classes (concepts) and properties (see http://wwww.holygoat.co.uk/owl/readwood/0.1/tags/): A tagger is encoded as an instance of *foaf:Agent*; the class *Holygoat:Tag* is defined as a subclass of *skos:Concept*; the property *Holygoat:relatedTag* is defined as a subproperty of *skos:semanticRelation*; and *Holygoat:taggedOn* is defined as a subproperty of *dct:date*. It is worth noting that, as with SIOC, the establishment of these relationships between the Holygoat Tag Ontology and external schemas supports simple inferences based on subsumption relations.

The MOAT ontology is characterized as a lightweight ontology that extends the Holygoat Tag Ontology by distinguishing between local meaning (i.e., the meaning of a tag assigned to a resource in a particular instance of tagging) and global meaning (i.e., all possible meanings of a tag in a particular system folksonomy) in order to represent the different meanings that are related to a single tag. MOAT assumes that there is a unique relationship between a tag and its meaning and that this tag-meaning relationship can be represented by a unique resource identifier (URI). Based on this premise, MOAT uses URIs to associate an instance of a tag to its intended semantic meaning.

## 4. Upper Tag Ontology (UTO)

The Upper Tag Ontology (UTO) is an upper-level ontology for social tagging that is designed to circumvent the complexity and potential redundancy inherent in user-generated tagging vocabularies. UTO is based on Gruber's (2007) suggestion that an ontology can be used to model tagging data, but it extends this idea with its focus on alignment between ontologies and the integration of tagging data with other sources of social metadata. UTO is not generally concerned with a Tagcloud, which is the central concept in SCOT; and it is defined in such a way that it can be aligned with other social metadata and tagging schemes, including FOAF, SIOC, SKOS and DCT. UTO emphasizes the structure of tagging behaviours rather than the meaning of the tags themselves, which distinguishes it from the MOAT ontology. By focusing on the structure of social tagging behaviours rather than tag semantics, this simple ontology can integrate metadata from one social tagging application with metadata from other social tagging Websites.

The Upper Tag Ontology (UTO) is defined as follows:

Let O be the UTO ontology, $\mathrm{O} = (C, \Re)$ (1)

Where $C = \{c_i, i \in N\}$ is a finite set of concepts and

$\Re = \{(c_i, c_k), i, k \in N\}$ is a finite set of relations established among the concepts in C.

In UTO, $C = \{Tag, Tagging, Object, Tagger, Source, Date, Comment, Vote\}$,

$$\Re = \left\{ \begin{matrix} hasRelatedTag, hasTag, hasObject, hasSource, hasDate, hasCreator, \\ hasComment, hasVote \end{matrix} \right\}$$

As expresssed in Definition (1) and illustrated in Figure 1, UTO consists of the eight concepts (or classes) described in Table 1 and the eight relations (or properties) described in Table 2. The *uto:Tagging* concept acts as a virtual node, in that it does not have a specific meaning but functions to gather or link the concepts relevant to a specific instance of tagging behaviour. Because a comment can be interpreted as having been added to the tag or to the object itself, many of the relations in UTO are defined as transitive. For example, an instance of *uto:Comment* can thus be connected to *uto:Object* or to *uto:Tag* via *uto:Tagging*. Thus, according to Definition (1), when $r \in \Re$, $i \in \mathrm{I}$ ($\mathrm{I}$ is the instances of ontology $\mathrm{O}$), $h, j, k \in N$

$r'$ is the inverse relation of $r$, when $i_j, i_k \in \mathrm{I}$, then $r(i_j) = i_k \Rightarrow r'(i_k) = i_j$

$r$ is transitive, when $i_h, i_j, i_k \in \mathrm{I}$, then $r(i_h) = i_j, r(i_j) = i_k \Rightarrow r(i_h) = i_k$

$r$ is symmetric, when $i_j, i_k \in \mathrm{I}$, then $r(i_j) = i_k \Leftrightarrow r(i_k) = i_j$

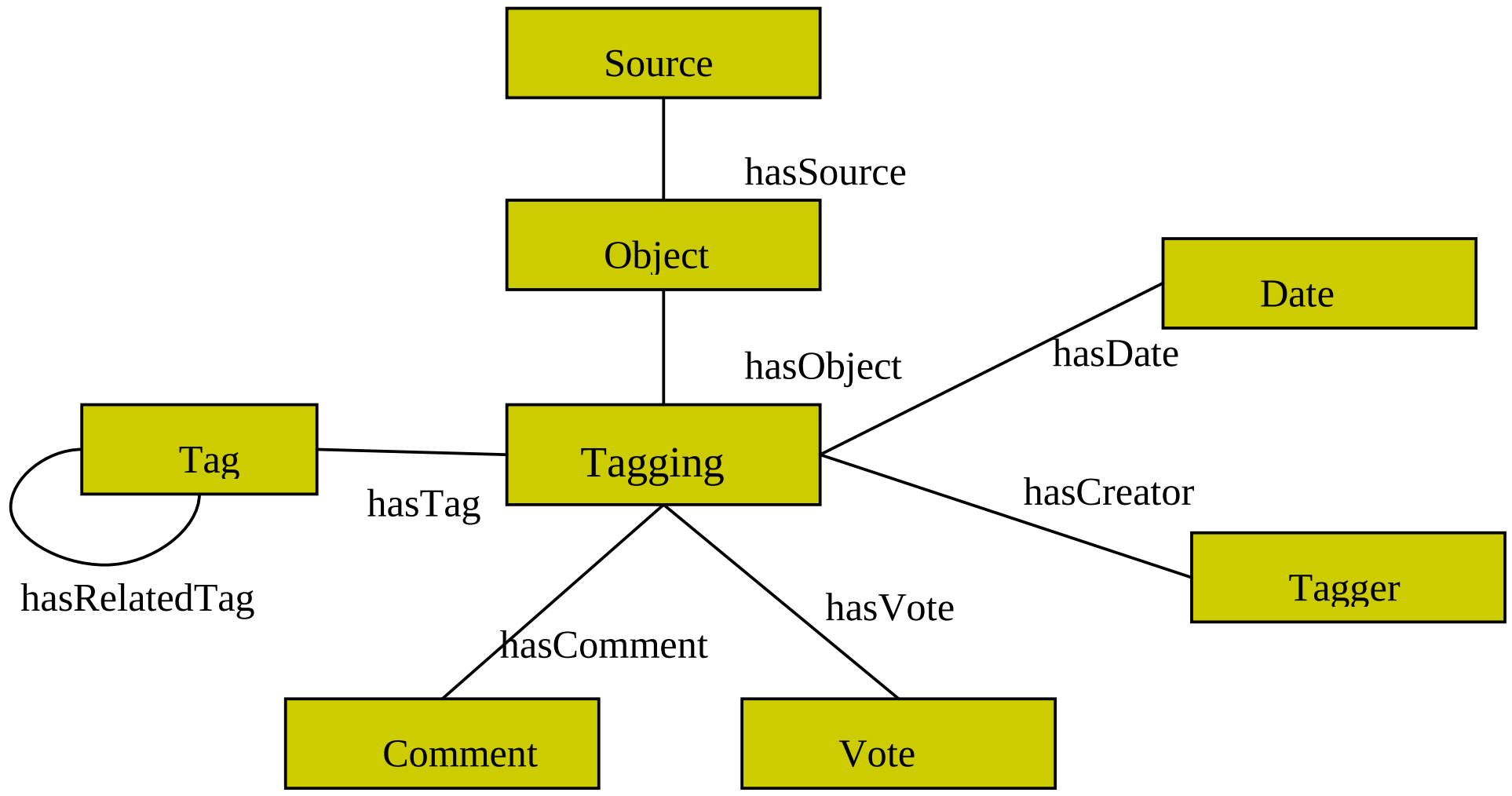

**Figure 1. The Upper Tag Ontology (UTO)**

**Table 1. Concepts in UTO**

| Concept | Synonyms | Description | Value Type | Instances |
|---|---|---|---|---|
| Tagging | | Tagging is the concept created to link other concepts. In itself, it does not have any real meaning. | | |
| Tag | Keyword | A Tag is a keyword that a Tagger assigns to an Object. | string | design, Web2.0, instructional_design, tutorials, etc. |
| Tagger | User | A Tagger is a userID for the person who assigns a Tag to an Object. | string | sborrelli |
| Object | Online object, Resource | An Object is an entity to which a Tagger assigns a Tag. It can be a bookmark (URL), photo, video, etc. | string | www.commoncraft.com/show |
| Source | Social network | A Source is an online site where the Tag-Object relationship is hosted. | string | Delicious, Flickr, YouTube, etc. |
| Comment | Note | A Comment is a statement added to an Object or Tag by a Tagger during the activity of Tagging. | string | The CommonCraft Show 1 Common Craft – Social Design for the Web |
| Date | Time | A Date is the time stamp of the Tagging activity. Format is "Mmm YY." | date | Jun 07 |
| Vote | Favourite | Tagging can be viewed as an act of voting. A Vote can be the number of different Taggers who assign a Tag to a bookmark in Delicious, a photo indicated as a favourite in Flickr, or the number of stars given to a video in YouTube. | integer | 103 (i.e., 103 taggers tagged this bookmark in Delicious) |

**Table 2. Relations in UTO**

| Relation | Domain | Range | Cardinality | OWL Type | Math properties | Inverse relation |
|---|---|---|---|---|---|---|
| hasTag | Object | Tag | -- | Object Property | Transitive | is_tag_of |
| hasRelatedTag | Tag | Tag | -- | Object Property | Transitive, Symmetric | -- |
| hasCreator | Tagging | Tagger | 1 | Object Property | Transitive | is_creator_of |
| hasObject | Tagging | Object | 1 | Object Property | -- | is_object_of |
| hasDate | Tagging | Date | 1 | Object Property | Transitive | -- |
| hasSource | Object | Source | -- | Object Property | -- | is_source_of |
| hasComment | Tagging | Comment | -- | Object Property | Transitive | is_comment_of |
| hasVote | Tagging | Vote | -- | Object Property | Transitive | is_vote_of |

While other ontologies developed to represent folksonomies tend to focus on the meanings of tags, UTO is designed to capture the structure of social tagging behaviour rather than the topic of a resource or the meaning of a tag. This shift in emphasis is intended to model the structure of the tagging data in order to integrate data from different tagging systems or to link distributed tagging records in order to effectively search or merge tagging data from different applications.

To avoid the generation of new problems and to mitigate the increase in complexity for applications, alignment or interoperability across ontologies should be as simple as possible. For this reason, alignment of UTO with other ontologies focuses primarily on mapping elements, both concepts/classes and relations/properties, using the relationship of either equivalence or subordination (see Table 3). For example, *uto:Tagger* is equivalent to *foaf:Person* and *sioc:User* and a subclass of *foaf:Agent* and *sioc:Usergroup*; *uto:Tag* is equivalent to *skos:Concept*; and *uto:Object* is defined as a superclass of *foaf:Document*, *foaf:Image* and *sioc:Post* using *owl:unionOf*. The relation *uto:hasRelatedTag* is equivalent to the properties *sioc:related_to* and *skos:narrower*, *skos:broader* and *skos:related*. Table 3 shows details of the alignment between the UTO ontology and FOAF, SIOC, DCT and SKOS. Appendix B provides the RDF/OWL definition of UTO and indicates the alignment of UTO elements with the elements of related social metadata schemes.

**Table 3. Ontology Alignment with UTO**

| UTO | FOAF | SIOC | DCT | SKOS |
| --- | --- | --- | --- | --- |
| Tagging | -- | -- | -- | -- |
| Tag | -- | -- | -- | = Concept |
| Tagger | = Person<br>⊆ Agent | = User<br>⊆ Usergroup | -- | -- |
| Object | ⊇Document<br>⊇Image | ⊇Post | -- | -- |
| Source | -- | ⊆ Community | -- | -- |
| Comment | -- | -- | -- | -- |
| Date | -- | -- | -- | -- |
| Vote | -- | -- | -- | -- |
| hasRelatedTag | -- | = related_to | -- | = narrower<br>= broader<br>= related |
| hasObject | -- | -- | -- | -- |
| hasSource | -- | = host_of | = source | -- |
| hasTag | = depiction<br>= topic | -- | -- | -- |
| hasVote | -- | -- | -- | -- |
| hasDate | -- | -- | -- | -- |
| hasCreator | = maker | = has_creator | = creator | -- |
| hasComment | -- | = note | -- | -- |

Note: According to Definition (1), $c_i, c_j \in C$, $c_i \subseteq c_j \Leftrightarrow c_i$ is the subclass of $c_j$, while, $c_i \supseteq c_j \Leftrightarrow c_i$ is the superclass of $c_j$, and $c_i = c_j \Leftrightarrow c_i$ is equivalent to $c_j$. These relationships are also valid for properties.

Aligning UTO with the social semantics of existing ontologies enables easy data integration, makes a mash-up of different semantics possible, and supports the linking of structured data. Using integrated data, tag searches can be performed across multiple sites and applications, and sources and relations (associations) can be mined across different platforms and applications. For example, using data linked through UTO, it is possible to find friends of Stefan who used the tag *spicy-Chinese-food* by aligning FOAF with UTO or to identify blogs, wikis or discussion groups where Stefan's friends have discussed "spicy Chinese food" by aligning FOAF and SOIC through UTO. Associations among tags, taggers and objects can also be mined. For example, networks of taggers can be mined through *foaf:knows* by aligning FOAF with UTO; relationships among tags can be mined with *skos:broader*, *skos:narrower* or *skos:related*; and co-occurrence technologies can be employed to mine associations among tags, taggers and objects. Following is an example of alignment between UTO and FOAF:

```
@prefix rdf: <http://www.w3.org/1999/02/22-rdf-syntax-ns#> .
@prefix uto: <http://info.slis.indiana.edu/~dingying/uto.owl#> .
@prefix foaf: <http://xmlns.com/foaf/0.1/> .
< http://info.slis.indiana.edu/~dingying/10357fc9-f6d2-4347-998c-aa26d63ef81b>
  uto:hasTag <http://del.icio.us/tag/social_networking> ;
  uto:hasVote "103" ;
  foaf:person "sborrellj" ;
  uto:hasObject <http://www.commoncraft.com/show>
  uto:hasComment "The CommonCraft Show | Common Craft - Social Design for the Web" ;
  uto:hasTag <http://del.icio.us/tag/design> ;
  uto:hasTag <http://del.icio.us/tag/instructional_design> ;
  uto:hasTag <http://del.icio.us/tag/tutorials> .
```

## 5. The UTO Crawler

To integrate tagging data from different social networks, we developed a tag crawler based on the Upper Tag Ontology (UTO) to harvest tagging data from Delicious, Flickr and YouTube and to store the retrieved data in RDF triples. To avoid timeouts and to make efficient use of available internet bandwidth, the UTO crawler uses the Smart and Simple Webcrawler framework, a multi-thread crawler designed by Torunsky (2008). There are two different parsers in the UTO crawler: one parses a page and searches for links that should be visited or filtered out, while the other parses HTML code to retrieve data about tags in accord with UTO. In general, the crawler collects data from the HTML coding and populates the elements of UTO accordingly. For example, when the crawler reaches a Webpage that contains tag data, it sends the information to Jena (http://jena.sourceforge.net/), which stores the data according to the UTO.

The UTO crawler contains six major components (see Figure 2). The main component acts as a host to initiate other components. The model component is responsible for the link logic that records already visited links. The filter component evaluates links and indicates which should be visited and which should be ignored. The crawler component coordinates executed tasks and distributes them to multiple threads. The parser component extracts the tagging data from HTML codes. And, finally, the RDF store component uses Jena to store extracted tagging data according to the classes and properties of the UTO.

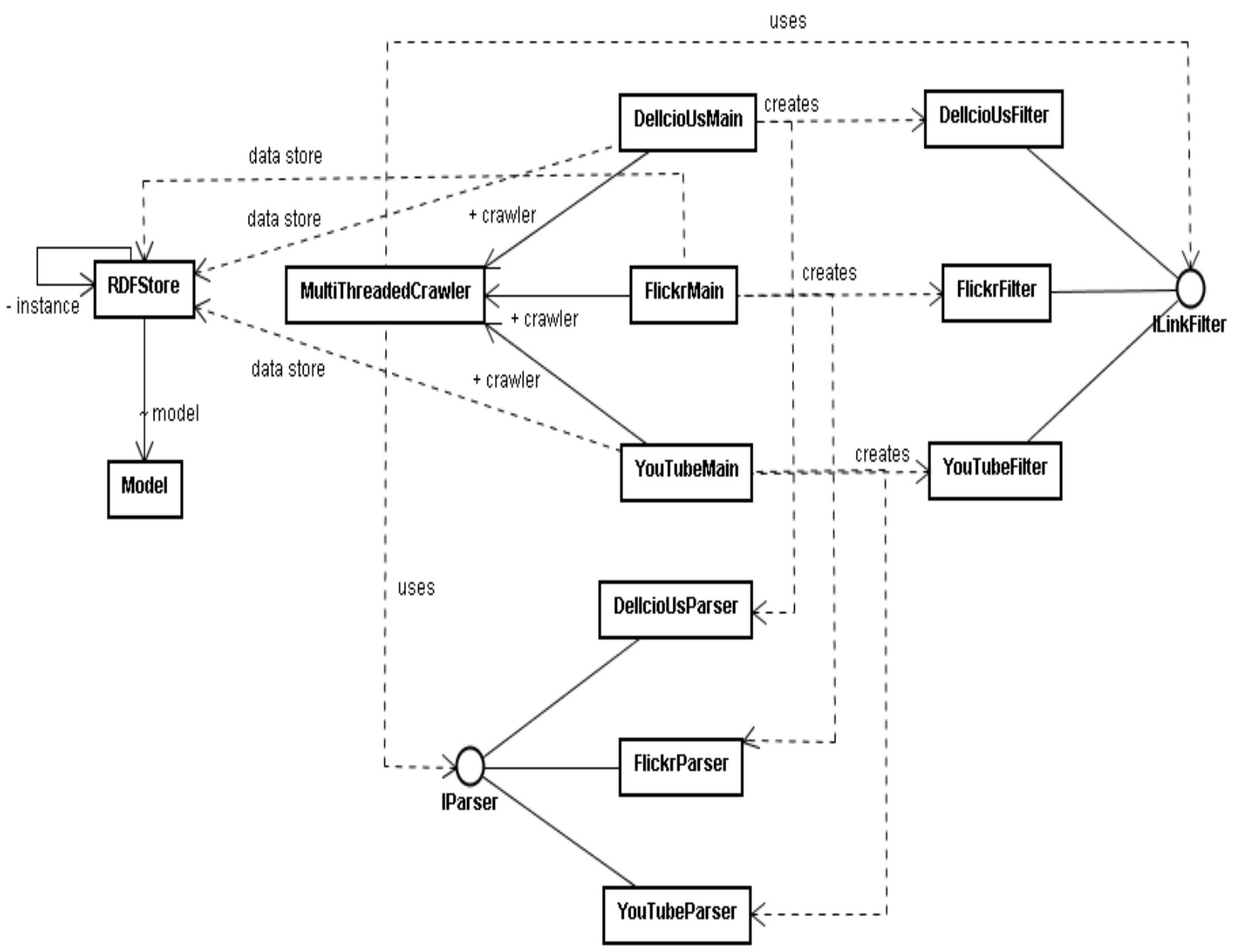

**Figure 2. Overview of UTO Crawler**

In Delicious, the crawler began with the tag cloud at http://delicious.com/tag and visited every tag in the cloud. For example, for TagA, the crawler visited the first Web page for TagA and parsed the HTML code to harvest information about bookmarks, taggers and related tags. For links bookmarked by only one tagger, tagging information was extracted from the TagA page. For bookmarks tagged by more than one tagger, the crawler went to the Delicious url for the bookmark and crawled the history of the bookmark to harvest information about which users had tagged this bookmark on which date(s). After gathering data about all of the bookmarks on the first page for TagA, the crawler continued to visit all remaining pages for TagA, performing the same tasks.

In Flickr, the crawler started from the tag cloud at http://flickr.com/photos/tags and visited each tag. For each tag page in Flickr, each photograph on the page was visited and information about the photograph, tags and tagger was extracted. In YouTube, the crawler started from the main page at http://youtube.com and visited all available video pages. For each video page, the crawler collected tagging data and visited all links pointing to other video pages. In order to avoid visiting the same page more than once, query parts of links were ignored. Table 4 provides examples of the data harvested from Delicious, Flickr and YouTube.

**Table 4. Values of the UTO Crawler**

| UTO | Delicious | Flickr | YouTube |
|---|---|---|---|
| hasTag | tag (e.g. http://delicious.com/tag/fashion) | tag (e.g. http://flickr.com/photos/tags/2006) | tag (e.g. http://www.youtube.com/results?search_query=Frenzy&search=tag) |
| hasRelatedTag | Supported | Supported | Not supported |
| hasObject | bookmark (e.g. http://www.twenty8twelve.com/) | photo url (e.g. http://farm3.static.flickr.com/2113/1899425530_7f10c0338a.jpg?v=0) | video url (e.g. http://www.youtube.com/watch?v=5ynmiKv2GcY) |
| hasCreator | tagger name (e.g. inna) | tagger name (e.g. rmen) | tagger name (e.g. karenyan119) |
| hasDate | date tagged (formatted mmm yy, e.g., Jun 07) | date tagged (formatted mmm yy, e.g. May 08) | date tagged (formatted mmm yy, e.g., Jan 08) |
| hasComment | statement about bookmark (not include title of bookmark | composed of photo title and statement from of a tagger | composed of video title and statement from a tagger |
| hasVote | number of different users tagging one bookmark | number of people favouring one photo | number of stars (1 to 5) assigned by a tagger |
| hasSource | http://delicious.com | photo page url (e.g. http://flickr.com/photos/samthesham/1991811650/) | http://youtube.com |

In September 2007, The UTO crawler was used to retrieve tagging data from Delicious, Flickr and YouTube. The crawler identified objects, taggers, tags, dates, comments and votes. In total, the data retrieved contains approximately 1 million bookmarks, 2.8 million taggers and 9.3 million tags harvested from Delicious; 296,000 photos, 154,000 taggers and 1.4 million tags harvested from Flickr; and 528,000 videos, 186,000 taggers and 1.4 million tags harvested from YouTube. The average number of tags per object ranges from 2.74 in YouTube to 9.31 in Delicious; the average number of tags assigned by a tagger ranges from 3.33 in Delicious to 8.79 in Flickr; and the average number of objects tagged by each tagger ranges from 0.36 in Delicious to 2.84 in YouTube. Although the average number of objects tagged by a user in Delicious seems abnormally low, this is a product of the Delicious bookmarking system: When a user uploads a bookmark to Delicious, he is required to provide a title for the URL, but he is not required to provide a tag. For this reason, Delicious has a number of bookmarks that have titles but no tags. Table 5 shows the details of the data harvested from each of the three social tagging sites.

**Table5. Crawled Tag Data**

| Social Network | Objects | Taggers | Tags | Tag/Object | Tag/Tagger | Object/Tagger |
|---|---|---|---|---|---|---|
| Delicious | 996,748 | 2,787,860 | 9,282,058 | 9.31 | 3.33 | 0.36 |
| Flickr | 295,837 | 153,778 | 1,351,201 | 4.57 | 8.79 | 1.92 |
| YouTube | 527,924 | 185,975 | 1,443,924 | 2.74 | 7.76 | 2.84 |

## 6. Integrating and Sb earching Tagging Data

One example of the tag data harvested from Delicious is represented below in both RDF/XML and Triple Turtle notation. In this example, one user tagged the resource http://www.commoncraft.com/show in June 2007. This bookmark was also tagged in Delicious by 103 other people using the tags *social_networking*, *design*, *Web2.0*, *instructional_design* and *tutorials*. The first tagger added the comment “The CommonCraft Show | Common Craft - Social Design for the Web.” Because the entry was crawled via the http://del.icio.us/tag/Web2.0 page, the tags *social_networking*, *design*,

*instructional_design* and *tutorials* are stored as tags related to *Web2.0*. The output tag data represented in UTO is shown using both RDF/XML and triple notation in Turtle syntax:

RDF/XML Notation:

```
<?xml version="1.0"?>
<rdf:RDF
        xmlns:rdf="http://www.w3.org/1999/02/22-rdf-syntax-ns#"
        xmlns:uto="http://info.slis.indiana.edu/~dingying/uto.owl#" >
   <rdf:Description rdf:about="http://info.slis.indiana.edu/~dingying/10357fc9-f6d2-4347-998c-aa26d63ef81b">
    <uto:hasTag rdf:resource="http://del.icio.us/tag/social_networking"/>
    <uto:hasVote>103</uto:hasVote>
    <uto:hasCreator>sborrelli</uto:hasCreator>
    <uto:hasObject rdf:resource="http://www.commoncraft.com/show"/>
    <uto:hasComment>The CommonCraft Show | Common Craft - Social Design for the Web</uto :hasComment>
    <uto:hasTag rdf:resource="http://del.icio.us/tag/design"/>
    <uto:hasDate>Jun 07</ uto:hasDate>
    <uto:hasTag rdf:resource="http://del.icio.us/tag/Web2.0"/>
    <uto:hasTag rdf:resource="http://del.icio.us/tag/instructional_design"/>
    <uto:hasTag rdf:resource="http://del.icio.us/tag/tutorials"/>
  </rdf:Description>
<rdf:RDF>
```

Triple Notation (Turtle format):

```
@prefix rdf: <http://www.w3.org/1999/02/22-rdf-syntax-ns#> .
@prefix uto: <http://info.slis.indiana.edu/~dingying/uto.owl#> .
< http://info.slis.indiana.edu/~dingying/10357fc9-f6d2-4347-998c-aa26d63ef81b>
   uto:hasTag <http://del.icio.us/tag/social_networking> ;
   uto:hasVote “103” ;
   uto:hasCreator “sborrellj” ;
   uto:hasObject <http://www.commoncraft.com/show> ;
   uto:hasComment “The CommonCraft Show | Common Craft - Social Design for the Web” ;
   uto:hasTag <http://del.icio.us/tag/design> ;
   uto;hasTag <http://del.icio.us/tag/instructional_design> ;
   uto:hasTag <http://del.icio.us/tag/tutorials> .
```

This example can also be represented as an RDF graph (see Figure 3).

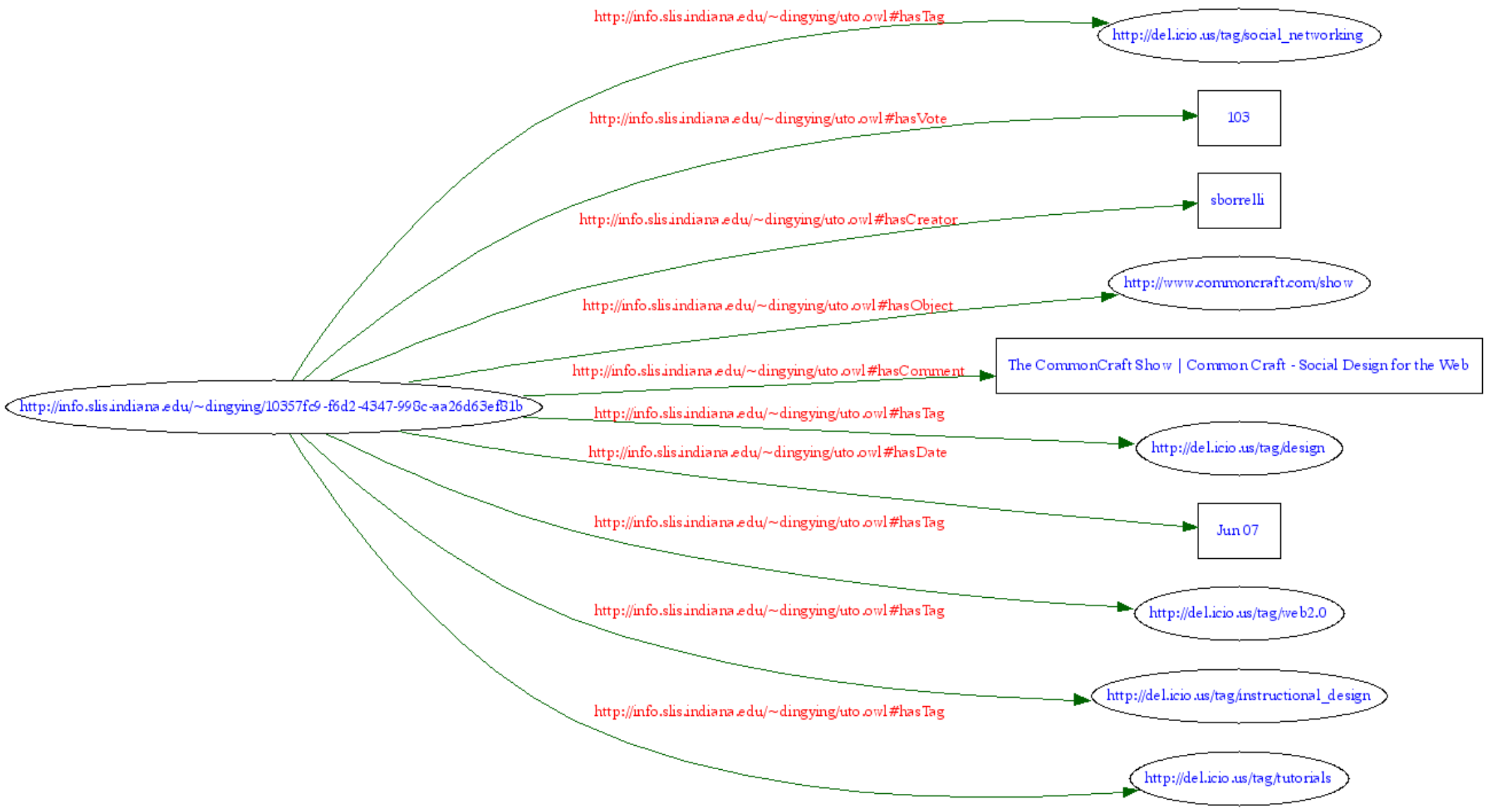

**Figure 3. RDF graph**

The tagging data from the Delicious, Flickr and YouTube websites was integrated according to UTO and stored in Jena. Based on this integration, some interesting queries can be performed. Three scenarios are presented here to demonstrate the possibilities for searching tagging data across the Delicious, Flickr and YouTube sites. The first scenario takes one tag as input and returns a list of the objects that have been assigned the same tag and their votes in descending order by vote (see Figure 4).

SPARQL query:

```
PREFIX xsd: <http://www.w3.org/2001/XMLSchema#>
select distinct ?object ?vote where {
{
?x <http://info.slis.indiana.edu/~dingying/uto.owl#hasObject> ?object .
?x <http://info.slis.indiana.edu/~dingying/uto.owl#hasVote> ?vote .
?x <http://info.slis.indiana.edu/~dingying/uto.owl#hasTag> <http://del.icio.us/tag/" + tag_text.getText() + ">
}
UNION
{
?x <http://info.slis.indiana.edu/~dingying/uto.owl#hasObject> ?object .
?x <http://info.slis.indiana.edu/~dingying/uto.owl#hasVote> ?vote .
?x <http://info.slis.indiana.edu/~dingying/uto.owl#hasTag> <http://flickr.com/photos/tags/" + tag_text.getText() + ">
}
UNION
{
?x <http://info.slis.indiana.edu/~dingying/uto.owl#hasObject> ?object .
?x <http://info.slis.indiana.edu/~dingying/uto.owl#hasVote> ?vote .
```

```
?x <http://info.slis.indiana.edu/~dingying/uto.owl#hasTag><http://youtube.com/results?search_query=" +
tag_text.getText() + "&amp;search=tag>
}
}order by desc(xsd:integer(?vote))
```

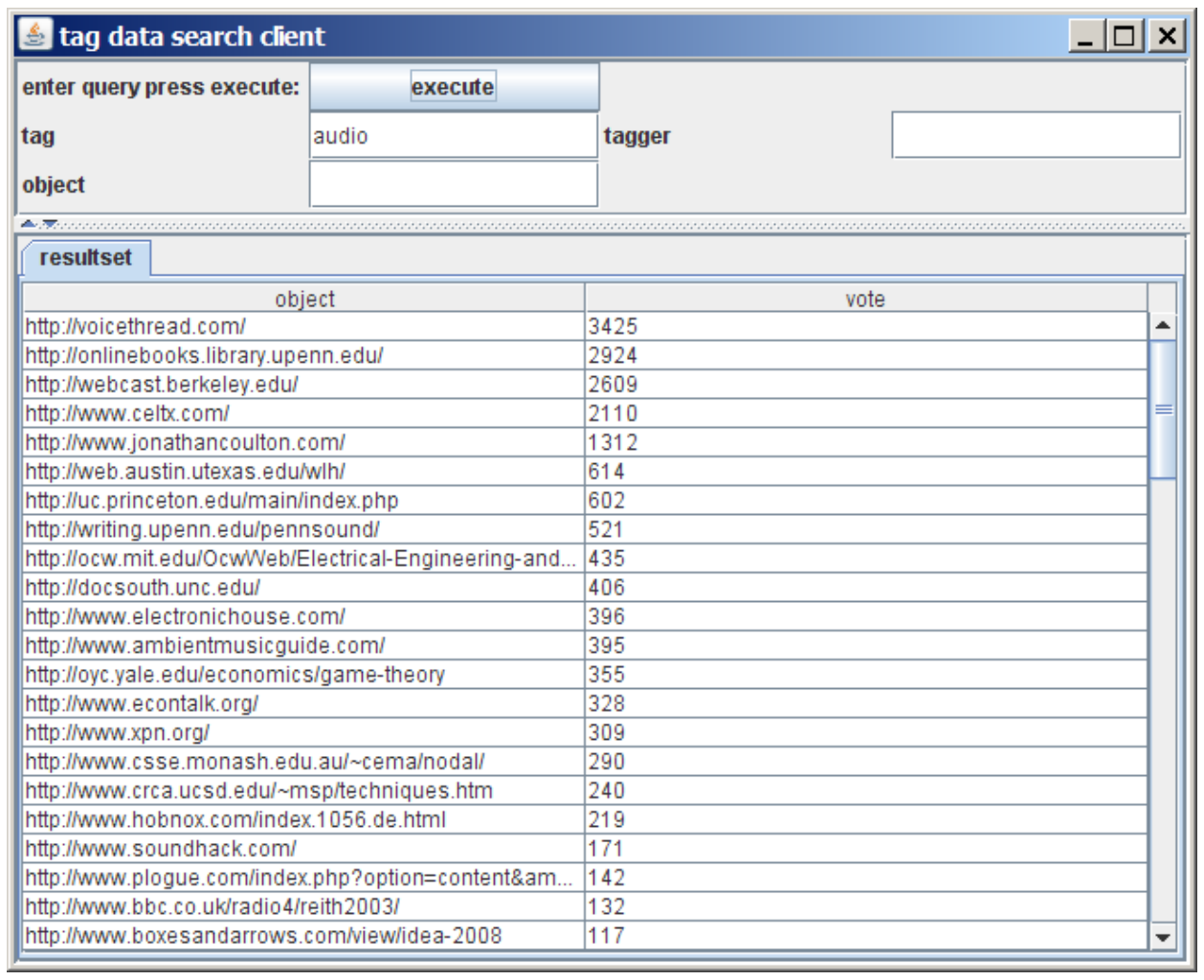

**Figure 4. Scenario 1 search frame**

The second scenario takes a single object as input and returns a list of taggers and tags for this object ordered alphabetically by tagger (see Figure 5).

SPARQL query:

```
select ?tagger ?tag where {
?x <http://info.slis.indiana.edu/~dingying/uto.owl#hasObject> <" + object_text.getText() + "> .
?x <http://info.slis.indiana.edu/~dingying/uto.owl#hasTag> ?tag .
?x <http://info.slis.indiana.edu/~dingying/uto.owl#hasCreator>  ?tagger
}
```

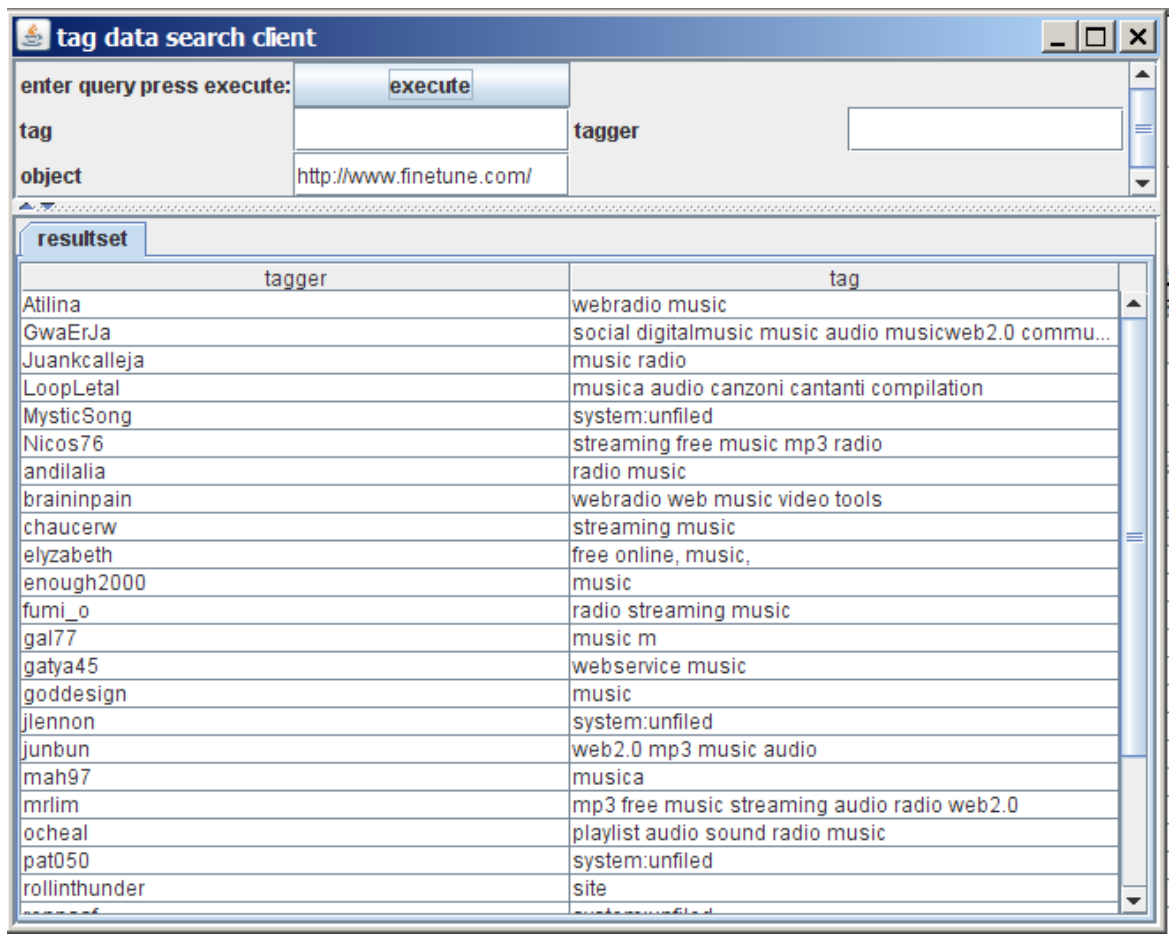

**Figure 5. Scenario 2 search frame**

The third scenario takes an individual tagger as input and returns a list of the objects tagged by that individual and the tags assigned to each object ordered alphabetically by object (see Figure 6).

SPARQL query:

```
select ?object ?tag where {
?x <http://info.slis.indiana.edu/~dingying/uto.owl#hasObject> ?object .
?x <http://info.slis.indiana.edu/~dingying/uto.owl#hasTag> ?tag .
?x <http://info.slis.indiana.edu/~dingying/uto.owl#hasCreator> \"" + tagger_text.getText() + "\"
}order by ?object";
```

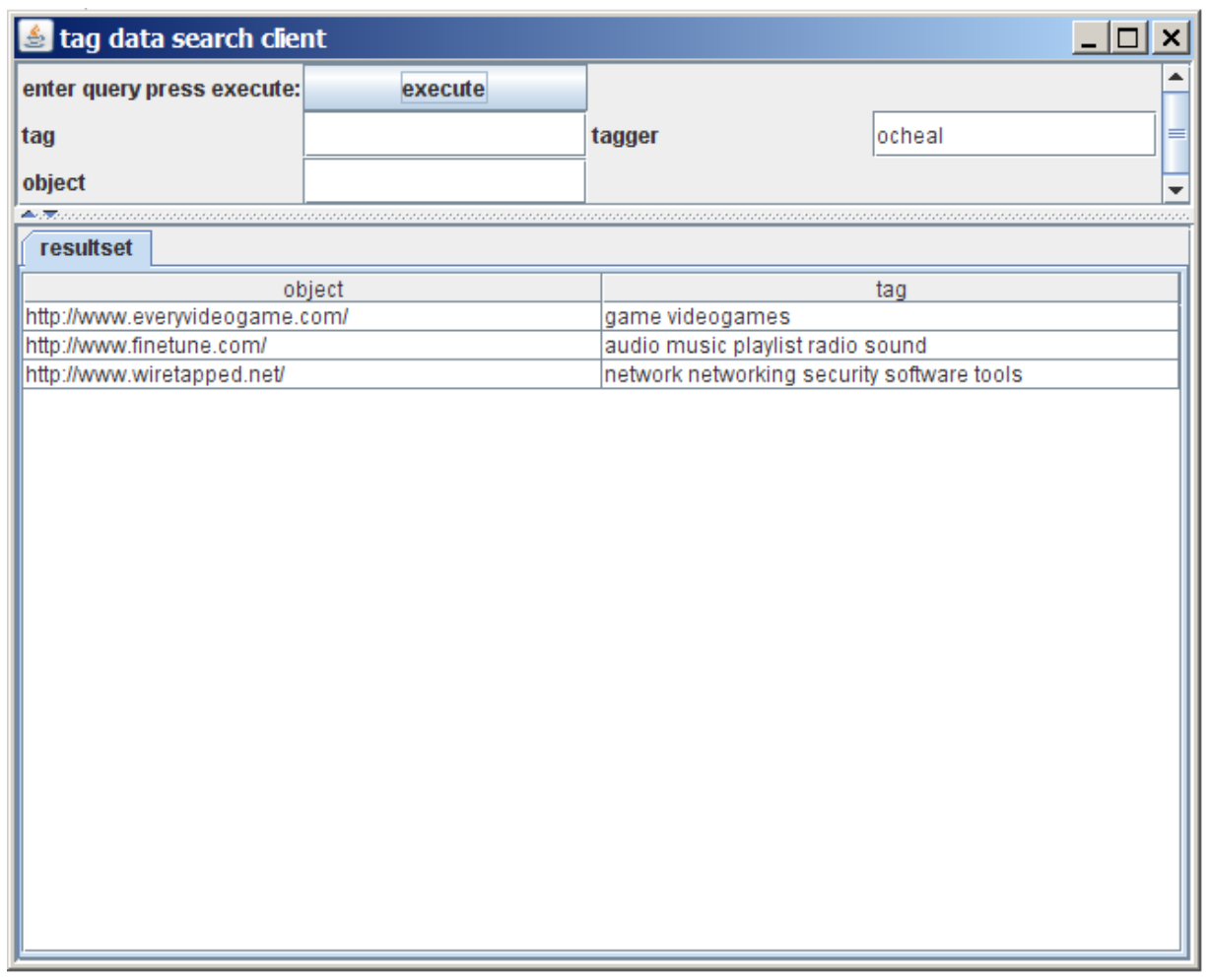

**Figure 6. Scenario 3 search frame**

## 7. Identifying Core Tag Set

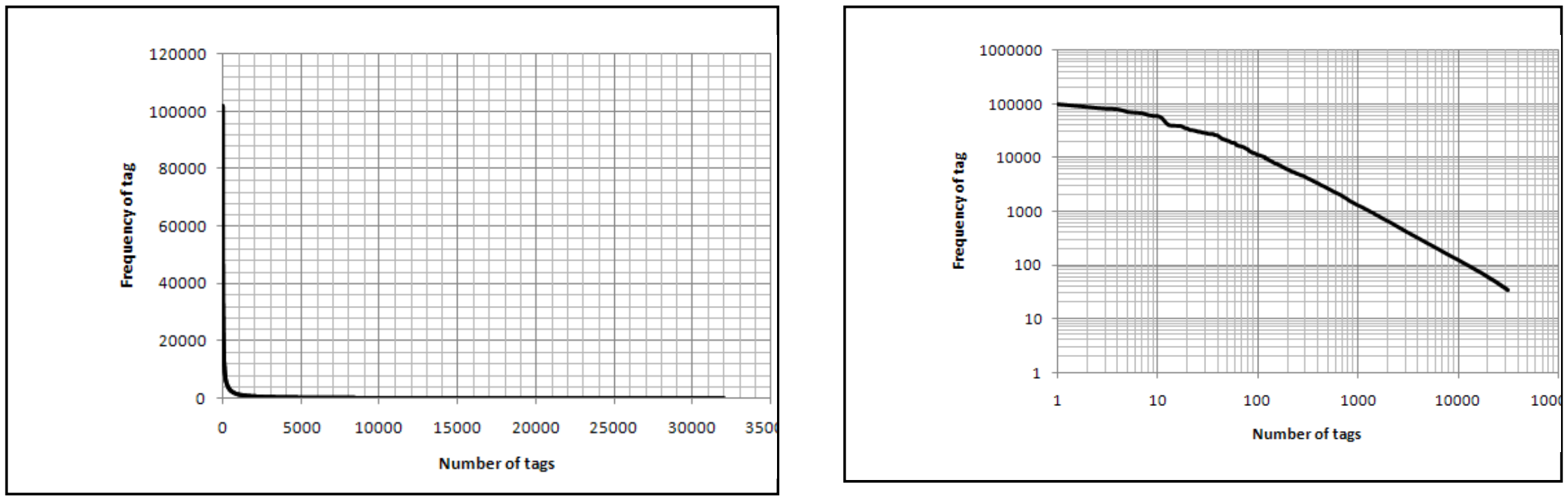

**Figure 7. Distribution of Tags**

Tagging data harvested from the three social networks -- Delicious, Flickr and YouTube -- were merged according to UTO to form a single, comprehensive dataset of tagging activities. Using this comprehensive dataset, we conducted a tag frequency analysis. Figure 7 demonstrates that the tag frequency distribution follows the power law distribution. Table 6 shows the details of this distribution: Only 1,363 of the 648,368 unique tags (or approximately 0.2% of all unique tags assigned between 2005 and September 2007) were assigned more than 1,000 times each, while 357,028 (or approximately 55% of all unique tags) were assigned only once. The results of this analysis are consistent with Zipf's Law, which states that a few tags will occur very often while many others will occur only rarely. The top 1,363 tags (see Appendix A) account for more than 50% of the total corpus of non-unique tags (see Table 6). *Design* is the most frequently occurring tag and accounts for 101,786 of 12,077,183 tags or nearly 1% of all tag occurrences. The second most frequently assigned tag is *blog* and accounts for 90,242 tag occurrences or 0.7% of all tag occurrences. Linguistic analysis of these 1,363 core tags will contribute to the identification of syntactic and semantic characteristics of the vocabularies generated through social tagging (Ding et al., submitted).

**Table 6. Tag Frequency Distribution**

| Tag Frequency Range | No. of unique tags | Cumulative % |
|---|---|---|
| 1 | 357028 | 55.07% |
| 2-10 | 217746 | 88.65% |
| 11-20 | 27404 | 92.88% |
| 21-30 | 11524 | 94.65% |
| 31-40 | 6656 | 95.68% |
| 41-50 | 4454 | 96.37% |
| 51-60 | 3387 | 96.89% |
| 61-70 | 2461 | 97.27% |
| 71-80 | 2066 | 97.59% |
| 81-90 | 1597 | 97.83% |
| 91-100 | 1348 | 98.04% |
| 101-200 | 6193 | 99.00% |
| 201-300 | 2151 | 99.33% |
| 301-400 | 1044 | 99.49% |
| 401-500 | 645 | 99.59% |
| 501-1,000 | 1301 | 99.79% |
| 1,001-120,000 | 1363 | 100.00% |

## 8. Conclusion

The Web is currently undergoing tremendous changes that bring with them significant challenges regarding the ability to connect information, knowledge, people and intelligence. Of the ongoing efforts to move Web 2.0 to the next level, work toward the Semantic Web is instilled with the long-term goal of fusing human and machine capabilities by representing data in machine-understandable ways and automating the mediation of data and services. As such, efforts to realize the Semantic Web are deeply embedded in the academic domain of artificial intelligence.

In the meantime, however, Web 2.0 has been successful in motivating users to collaborate with each other and to share information via the Web (Hinchcliffe, 2006). And Web 2.0 is not entirely different from the Semantic Web. As Tim Berners-Lee (2001) states, "The Semantic Web is an extension of the current Web in which information is given well-defined meaning, better enabling computers and people to work in cooperation." Web 2.0 not only extends the communication dimensions of publication, commentary and argument but also contributes contextual information to the current Web -- contextual information in the form of "social metadata" that has been generated informally by users through the tagging, bookmarking and annotating of online resources.

The power of the Semantic Web will reside in its potential to support interoperability through the development, deployment and application of well-defined metadata -- metadata that supports logical reasoning and is encoded in a machine-understandable language. The module and layer design principles of the Semantic Web (e.g., URIs, RDF/RDFS, ontologies, and logical languages, etc.) will pave the way for reuse of existing data and the introduction of intelligent (and more efficient) search facilities that will support greater granularity and higher relevance of result sets. While Web 2.0 provides a scalable, community-powered information sharing platform, the Semantic Web will add valuable machine-understandable metadata that enable efficient and automatic approaches to collaboration, cross-portal communication and the sharing of heterogeneous information.

Social aspects of the Web necessarily influence the sharing and use of information. The Social Web relies on users to identify and link useful content and to provide feedback. The participation of growing numbers of users has significantly increased both the heterogeneity and the trustworthiness of the Web; and it has created a social approach to data integration that capitalizes on collective intelligence. By tagging and sharing data, users create relationships between resources and enrich the contextual information associated with resources and concepts. The social tagging systems in which users participate also provide examples that identify pragmatic ways of using the phenomena of the Social Web to realize data mediation and integration. Thus the Upper Tag Ontology (UTO) is proposed to represent social tagging data, to integrate metadata from different tagging systems and to mediate between heterogeneous social metadata.

In the future, we hope to extend these ideas to the Social Web so as to integrate data based on collective intelligence through the consideration of instances and the contextual information provided by Web users. We also hope to mine associations among tagging data to discover hidden associations in complex social tagging networks using tag co-occurrence analysis -- measuring the co-occurrence of tags based either on objects or on taggers, for example -- in combination with advanced techniques for social network analysis -- macro-level network indicators such as scale-free, cluster coefficient, and k-core as well as micro-level network indicators such as degree, betweenness, closeness and eigenvector centralities.

## Acknowledgements

The authors would like to thank the University of Innsbruck, where data collection and analysis were conducted.

The authors are also very grateful to three anonymous reviewers for their insightful comments.

# Appendix A:

Core tag vocabulary in three social networks: Top 1363 tags in Delicious, Flickr and YouTube

| | Core Tag Vocabulary |
|---|---|
| Numbers & others | 1, 2, 3, 2005, 2006, 2007, -, .net, 3d |
| A | a, academia, academic, accessibility, accessories, acoustic, action, actionscript, activism, ad, admin, administration, adobe, ads, adsense, adult, advertising, advice, Africa, agency, aggregator, agile, ai, air, airline, airlines, airplane, airport, ajax, algorithm, algorithms, all, alternative, amateur, amazing, amazon, America, American, Amsterdam, analysis, analytics, and, angel, angst, animal, animals, animation, anime, anonymous, anthropology, apache, api, apple, application, applications, apps, architecture, archive, archives, argentina, art, arte, article, articles, artist, artists, arts, as3, asia, asian, asp.net, ass, asterisk, astronomy, at, atheism, atom, au, audio, audiobooks, Australia, authentication, auto, automation, autumn, awards, awesome |
| B | Baby, backup, bad, ball, band, bands, bandslash, bank, banking, bar, Barcelona, baseball, bass, bbc, beach, beatles, beautiful, beauty, beer, berlin, best, bible, bibliography, bicycle, big, bike, bioinformatics, biology, bird, birds, birthday, bit200f06, bit200w07, bittorrent, black, blackandwhite, blog, blogger, blogging, blogs, blood, blue, Bluetooth, boat, body, boobs, book, bookmarking, bookmarks, books, boston, boy, boys, bpm, brain, branding, brasil, brazil, bridge, Britney, Brooklyn, brown, browser, browsers, Buddhism, building, bus, bush, business, buy, bw, by |
| C | C, c#, c++, calculator, calendar, California, camera, cameraphone, camping, Canada, canon, car, card, cards, career, cars, cartoon, cartoons, cat, cats, cd, celebrity, cell, cellphone, celltagged, censorship, change, charity, charts, chat, cheap, cheatsheet, chemistry, Chicago, chicken, child, children, chile, china, Chinese, chocolate, chords, chris, Christian, Christianity, Christmas, church, ciencia, cine, cinema, city, class, classic, classification, climate, clip, clothes, clothing, clouds, club, cluster, clustering, cms, cocoa, code, coding, coffee, collaboration, collection, college, color, colors, colour, comedy, comic, comics, commercial, communication, community, company, comparison, competition, compiler, complexity, computer, computers, computing, concert, concurrency, conference, conferences, conspiracy, consumer, content, contest, control, conversion, convert, converter, cooking, cool, copyright, corporate, country, course, courses, cover, crack, craft, crafts, crazy, creative, creativecommons, creativity, credit, crime, crossover, cryptography, cs, css, cultura, culture, curiosidades, custom, cute, cycling |
| D | Daily, dance, dancing, dark, data, database, datamining, dating, david, day, dc, de, dead, deals, death, debian, del.icio.us, delicious, demo, democracy, design, designer, desktop, deutsch, Deutschland, dev, developer, development, dhtml, dictionary, diet, dig, digital, directory, diseÃ±o, Disney, distributed, distro, diy, dj, django, dns, do, documentary, documentation, dog, dogs, dom, domain, dotnet, download, downloads, drawing, driver, drm, drugs, drunk, drupal, duesouth, dvd |
| E | Earth, ebay, ebook, ebooks, eclipse, ecology, ecommerce, economia, economics, economy, editing, editor, edtech, educaciÃ³n, educacion, education, effects, el, elearning, e-learning, electronic, electronics, email, embedded, employment, emulation, en, encryption, encyclopedia, energy, engine, engineering, England, English, enterprise, enterprise2.0, entertainment, entrepreneur, entrepreneurship, environment, erlang, esl, espaÃ±a, espaÃ±ol, essay, ethics, eu, europa, Europe, event, events, evolution, examples, excel, exchange, exercise, experimental, extension, extensions, eyes |
| F | f1, face, facebook, fall, family, fanfic, fanfiction, fantasy, faq, fashion, fat, feed, feeds, female, feminism, festival, fetish, fic, fiction, fight, file, files, filesharing, filesystem, film, films, finance, financial, fire, firefox, firefox:bookmarks, firefox:rss, firefox:toolbar, firewall, fish, fitness, flash, flex, flickr, flight, flights, florida, flower, flowers, fob, folksonomy, font, fonts, food, football, for, forms, forum, forums, foto, fotografia, fotos, framework, france, free, freedom, freelance, freeware, French, friends, from, fuck, fun, functional, funny, furniture, future |
| G | Gadget, gadgets, gallery, game, games, gaming, garden, gardening, gay, gear, geek, gen, gender, genealogy, generator, genetics, geo, geography, George, geotagged, german, germany, ghost, gifts, girl, girls, gis, glass, global, gmail, gnome, gnu, go, god, good, google, googlemaps, government, gps, graffiti, grammar, graph, graphic, graphicdesign, graphics, gratis, great, green, grid, gtd, gui, guide, guitar |
| H | Hack, hacking, hacks, hair, Halloween, halo, happiness, happy, hardware, Haskell, hci, hdr, health, healthcare, heart, Hebrew, help, het, hibernate, high, hip, hiphop, history, holiday, home, hop, horror, hosting, hot, hotel, hotels, house, housing, how, howto, hp, html, http, human, humor, humour |
| I | I, ia, ibm, ical, icon, icons, ict, ide, idea, ideas, identity, ie, illustration, illustrator, im, image, images, imported, in, india, indie, info, informatica, information, innovation, inspiration, install, installation, insurance, intel, intelligence, interaction, interactive, interesting, interface, interior, international, internet, interview, investing, investment, ip, iphone, ipod, iptv, iran, Iraq, irc, Ireland, is, islam, island, Israel, it, italia, Italian, Italy, itunes |
| J | j2ee, jabber, jack, james, japan, Japanese, java, javascript, jazz, jesus, jewelry, job, jobs, john, joomla, journal, journalism, journals, jsf, json, juegos |
| K | Kernel, keyboard, kid, kids, king, kiss, knitting, knowledge, korea, korean |
| L | La, lake, landscape, language, languages, laptop, latex, latin, law, layout, learn, learning, leaves, lectures, legal, lego, lesbian, lessons, libraries, library, library2.0, libros, life, lifehack, lifehacker, lifehacks, lifestyle, light, lighting, lights, linguistics, link, links, linux, lisp, list, lists, literacy, literature, literature, little, live, local, logic, logo, lol, London, long, los, losangeles, love, lyrics |
| M | Mac, macbook, macintosh, macosx, macro, Madrid, magazine, magazines, magic, mail, make, man, management, manga, manual, mÃºsica, map, mapas, mapping, maps, market, marketing, mashup, math, mathematics, maths, mckay/Sheppard, me, media, medical, medicine, memory, men, menu, messaging, metadata, metal, mexico, Michael, microformats, Microsoft, midi, military, mind, mindmap, misc, mit, mix, mobile, model, modeling, models, modern, module, money, monitor, monitoring, motion, motiongraphics, motivation, mountain, movie, |
| | movies, Mozilla, mp3, multimedia, museum, music, Musica, musik, my, myspace, mysql |
| N | Naked, naruto, nasa, national, nature, navigation, nc-17, Netherlands, network, networking, networks, new, newmedia, news, newspaper, newspapers, newyork, night, Nikon, Nintendo, nlp, no, nokia, nonprofit, notes, noticias, November, nptech, nude, nutrition, nyc |
| O | Ocean, October, of, office, oil, old, on, one, online, ontology, open, opened, openoffice, opensource, open-source, opera, opinion, optimization, oracle, orange, organic, organization, origami, os, osx, out, outdoors, outlook, owl |
| P | p2p, painting, palm, paper, papers, parenting, paris, park, parody, parser, parsing, party, password, pattern, patterns, paul, pc, pda, pdf, peace, people, performance, perl, personal, personality, pet, pets, philosophy, phone, photo, photographer, photography, photos, photoshop, php, physics, piano, picture, pictures, pink, planning, plants, play, player, plugin, plugins, pocketpc, podcast, podcasting, podcasts, poetry, poker, Poland, police, policy, polish, politics, politik, pop, porn, portable, portal, portfolio, portrait, Portugal, post, power, powerpoint, pr, presentation, presentations, print, printing, privacy, process´, processing, product, production, productivity, products, programming, project, projectmanagement, projects, property, prototype, proxy, psychology, public, publishing, punk, puppy, pussy, puzzle, python |
| Q | quotes |
| R | r, race, racing, radio, rails, random, rap, rdf, read, reading, real, realestate, recherché, recipe, recipes, recording, records, recovery, recursos, red, reference, reflection, regex, religion, remix, remote, repair, research, resource, resources, rest, restaurant, restaurants, retro, review, reviews, rights, river, road, robot, robotics, robots, rock, roma, rome, rpg, rps, rss, ruby, rubyonrails, running, Russia, russian |
| S | Safari, safari_export, sam/dean, san, sanfrancisco, satellite, scary, scheme, school, science, scifi, Scotland, screen, script, scripting, scripts, sculpture, sea, search, searchengine, searchengines, seattle, secondlife, security, seguridad, self, semantic, semanticWeb, semWeb, seo, series, server, service, services, sewing, sex, sexy, sf, sga, share, sharepoint, sharing, shell, shoes, shop, shopping, short, show, simulation, singing, site, sky, skype, slash, sleep, slideshow, smallville, sms, snow, soa, soap, soccer, social, socialmedia, socialnetworking, socialnetworks, socialsoftware, society, sociology, software, solar, song, songs, sony, sound, source, south, space, spain, spam, Spanish, spears, speech, speed, spirituality, spn, sport, sports, spring, sql, ssh, standards, star, startup, starwars, statistics, stats, stock, stocks, storage, store, stories, story, strategy, streaming, street, streetart, studio, study, stuff, stupid, style, subversion, summer, sun, sunset, super, supernatural, support, sustainability, svn, Sweden, sweet, swing, Switzerland, symbian, sync, sysadmin, system |
| T | Tabs, tag, tagging, tags, Taiwan, teaching, tech, techno, technology, tecnologia, telephone, television, template, templates, terrorism, test, testing, texas, text, the, theme, themes, theory, thesis, time, tips, to, todo, Tokyo, tom, tool, tools, top, toread, Toronto, torrent, torrents, tour, tourism, toy, toys, trabajo, tracking, trading, traffic, trailer, train, training, translation, transport, transportation, travel, tree, trees, trends, tricks, trip, tuning, tutorial, tutorials, tutorials, tv, twitter, type, typography |
| U | Ubuntu, ui, uk, uml, uni, university, unix, unread, up, upload, urban, us, usa, usability, usb, useful, usenet, utilities, utility, ux |
| V | Vacation, validation, Vancouver, vc, vector, vegetarian, viajes, video, videogames, videos, vim, vintage, vinyl, viral, virtual, virtualization, vista, visual, visualization, vmware, voip, vs |
| W | w3c, wall, wallpaper, wallpapers, war, Washington, water, weather, Web, Web2.0, Webapp, Webcam, Webcomic, Webdesign, Webdev, Webdevelopment, Weblog, Webmaster, Webservice, Webservices, Website, Websites, Webstandards, Webtools, wedding, weird, white, widget, widgets, wifi, wii, wiki, Wikipedia, wikis, window, windows, wine, winter, wireless, wishlist, with, woman, women, wood, word, wordpress, words, work, workflow, world, wow, writing, wysiwyg |
| X | X, xbox, xhtml, xml, xp, xslt, xxx |
| Y | Yahoo, yellow, York, you, young, your, youth, youtube |
| Z | Zombie, zoo |

# Appendix B:

UTO in RDF/OWL format (uto.owl)

```
<?xml version="1.0"?>
<rdf:RDF
    xmlns:rdf="http://www.w3.org/1999/02/22-rdf-syntax-ns#"
    xmlns:xsd="http://www.w3.org/2001/XMLSchema#"
    xmlns:uto="http://info.slis.indiana.edu/~dingying/uto.owl#"
    xmlns:rdfs="http://www.w3.org/2000/01/rdf-schema#"
    xmlns:owl="http://www.w3.org/2002/07/owl#"
    xmlns:dct="http://dublincore.org/2008/01/14/dcterms.rdf#"
    xmlns:foaf="http://xmlns.com/foaf/0.1/#"
    xmlns:skos="http://www.w3.org/2004/02/skos/core#"
```

```
  xmlns:sioc="http://rdfs.org/sioc/ns#"
  xml:base="http://info.slis.indiana.edu/~dingying/uto.owl#">

<owl:Ontology rdf:about=""/>

<owl:Class rdf:about="http://info.slis.indiana.edu/~dingying/uto.owl#Tag">
  <rdfs:comment rdf:datatype="http://www.w3.org/2001/XMLSchema#string">A tag is a keyword that a user
    adds to an object.</rdfs:comment>
  <rdfs:label>Tag</rdfs:label>
  <owl:equivalentClass rdf:resource="http://www.w3.org/2004/02/skos/core#Concept"/>
</owl:Class>

<owl:Class rdf:about="http://info.slis.indiana.edu/~dingying/uto.owl#Comment">
  <rdfs:comment rdf:datatype="http://www.w3.org/2001/XMLSchema#string">A comment is the statement or
    set of statements that a tagger adds to an object or tag during the act of tagging.</rdfs:comment>
  <rdfs:label>Comment</rdfs:label>
</owl:Class>

<owl:Class rdf:about="http://info.slis.indiana.edu/~dingying/uto.owl#Source">
  <rdfs:comment rdf:datatype="http://www.w3.org/2001/XMLSchema#string">Source is the place where the
    object is hosted. It can be delicious, flickr, youtube, etc.</rdfs:comment>
  <rdfs:subClassOf rdf:resource="http://rdfs.org/sioc/ns#Community"/>
  <rdfs:label>Source</rdfs:label>
</owl:Class>

<owl:Class rdf:about="http://info.slis.indiana.edu/~dingying/uto.owl#Vote">
  <rdfs:comment rdf:datatype="http://www.w3.org/2001/XMLSchema#string">Tagging can be viewed as voting.
    Vote can be the number of different taggers tagging a bookmark, a photo or a video as
    favoriate.</rdfs:comment>
  <rdfs:label>Vote</rdfs:label>
</owl:Class>

<owl:Class rdf:about="http://info.slis.indiana.edu/~dingying/uto.owl#Date">
  <rdfs:comment rdf:datatype="http://www.w3.org/2001/XMLSchema#string">Date is the time stamp o fte
    tagging behavior. Format is "MmmYY"</rdfs:comment>
  <rdfs:label>Date</rdfs:label>
</owl:Class>

<owl:Class rdf:about="http://info.slis.indiana.edu/~dingying/uto.owl#Tagger">
  <rdfs:comment rdf:datatype="http://www.w3.org/2001/XMLSchema#string">Tagger is the user who tags
    object</rdfs:comment>
```

```
    <rdfs:label>Tagger</rdfs:label>
    <rdfs:subClassOf rdf:resource="http://xmlns.com/foaf/0.1/Agent"/>
    <rdfs:subClassOf rdf:resource="http://rdfs.org/sioc/ns#Usergroup"/>
    <owl:equivalentClass rdf:resource="http://xmlns.com/foaf/0.1/Person"/>
    <owl:equivalentClass rdf:resource="http://rdfs.org/sioc/ns#User"/>
</owl:Class>

<owl:Class rdf:about="http://info.slis.indiana.edu/~dingying/uto.owl#Tagging">
  <rdfs:comment rdf:datatype="http://www.w3.org/2001/XMLSchema#string">Tagging is the concept which is
    created to link other concepts. Itself does not have any real meaning</rdfs:comment>
  <rdfs:label>Tagging</rdfs:label>
</owl:Class>

<owl:Class rdf:about="http://info.slis.indiana.edu/~dingying/uto.owl#Object">
  <rdfs:comment rdf:datatype="http://www.w3.org/2001/XMLSchema#string">object is the thing which tagger
    is tagging. It can be bookmarks, photos, videos, musics, books, slides, etc.</rdfs:comment>
  <rdfs:label>Object</rdfs:label>
  <owl:unionOf rdf:parseType="Collection">
    <owl:Class rdf:about="http://xmlns.com/foaf/0.1/Document"/>
    <owl:Class rdf:about="http://xmlns.com/foaf/0.1/Image"/>
    <owl:Class rdf:about="http://www.w3.org/2000/01/rdf-schema#Resource"/>
    <owl:Class rdf:about="http://rdfs.org/sioc/ns#Post"/>
  </owl:unionOf>
</owl:Class>

<owl:ObjectProperty rdf:about="http://info.slis.indiana.edu/~dingying/uto.owl#hasDate">
  <rdfs:label>hasDate</rdfs:label>
  <rdfs:domain rdf:resource="#Tagging"/>
  <rdfs:range rdf:resource="#Date"/>
  <owl:equivalentProperty rdf:resource="http://purl.org/dc/elements/1.1/date"/>
</owl:ObjectProperty>

<owl:ObjectProperty rdf:about="http://info.slis.indiana.edu/~dingying/uto.owl#hasObject">
  <rdfs:label>hasObject</rdfs:label>
  <rdfs:range rdf:resource="#Object"/>
  <rdfs:domain rdf:resource="#Tagging"/>
</owl:ObjectProperty>

<owl:ObjectProperty rdf:about="http://info.slis.indiana.edu/~dingying/uto.owl#hasVote">
  <rdfs:label>hasVote</rdfs:label>
```

```
  <rdfs:range rdf:resource="#Vote"/>
  <rdfs:domain rdf:resource="#Tagging"/>
</owl:ObjectProperty>

<owl:ObjectProperty rdf:about="http://info.slis.indiana.edu/~dingying/uto.owl#hasTag">
  <rdfs:label>hasTag</rdfs:label>
  <rdfs:domain rdf:resource="#Tagging"/>
  <rdfs:range rdf:resource="#Tag"/>
  <owl:equivalentProperty rdf:resource="http://purl.org/dc/elements/1.1/description"/>
  <owl:equivalentProperty rdf:resource="http://xmlns.com/foaf/0.1/depiction"/>
  <owl:equivalentProperty rdf:resource="http://xmlns.com/foaf/0.1/topic"/>
</owl:ObjectProperty>

<owl:ObjectProperty rdf:about="http://info.slis.indiana.edu/~dingying/uto.owl#hasComment">
  <rdfs:label>hasComment</rdfs:label>
  <rdfs:range rdf:resource="#Comment"/>
  <rdfs:domain rdf:resource="#Tagging"/>
  <owl:equivalentProperty rdf:resource="http://rdfs.org/sioc/ns#note"/>
</owl:ObjectProperty>

<owl:ObjectProperty rdf:about="http://info.slis.indiana.edu/~dingying/uto.owl#hasRelatedTag">
  <rdfs:label>hasRelatedTag</rdfs:label>
  <rdfs:domain rdf:resource="#Tag"/>
  <rdfs:range rdf:resource="#Tag"/>
  <owl:equivalentProperty rdf:resource="http://rdfs.org/sioc/ns#related_to"/>
  <owl:equivalentProperty rdf:resource="http://www.w3.org/2004/02/skos/core#broader"/>
  <owl:equivalentProperty rdf:resource="http://www.w3.org/2004/02/skos/core#narrower"/>
  <owl:equivalentProperty rdf:resource="http://www.w3.org/2004/02/skos/core#related"/>
</owl:ObjectProperty>

<owl:ObjectProperty rdf:about="http://info.slis.indiana.edu/~dingying/uto.owl#hasCreator">
   <rdfs:label>hasCreator</rdfs:label>
  <rdfs:range rdf:resource="#Tagger"/>
  <rdfs:domain rdf:resource="#Tagging"/>
  <owl:equivalentProperty rdf:resource="http://purl.org/dc/elements/1.1/creator"/>
  <owl:equivalentProperty rdf:resource="http://xmlns.com/foaf/0.1/maker"/>
  <owl:equivalentProperty rdf:resource="http://rdfs.org/sioc/ns#has_creator"/>
</owl:ObjectProperty>
```

```
  <owl:InverseFunctionalProperty rdf:about="http://info.slis.indiana.edu/~dingying/uto.owl#hasSource">
    <rdfs:label>hasSource</rdfs:label>
    <rdfs:range rdf:resource="#Source"/>
    <rdfs:domain rdf:resource="#Object"/>
    <rdf:type rdf:resource="http://www.w3.org/2002/07/owl#ObjectProperty"/>
    <owl:equivalentProperty rdf:resource="http://rdfs.org/sioc/ns#host_of"/>
    <owl:equivalentProperty rdf:resource="http://purl.org/dc/elements/1.1/source"/>
  </owl:InverseFunctionalProperty>

</rdf:RDF>
```

Notes: uot.owl has been validated via OWL Validator at http://www.mygrid.org.uk/OWL/Validator